\documentclass[10pt]{article}

\usepackage[utf8]{inputenc} 
\usepackage[T1]{fontenc}
\usepackage[english]{babel}
\usepackage{csquotes}
\usepackage[cmex10]{amsmath}  
\usepackage{amsfonts}
\usepackage{amssymb}
\usepackage[a4paper]{geometry}

\newtheorem{theorem}{\textbf{Theorem}}[section]
\newtheorem{proposition}[theorem]{\textbf{Proposition}}
\newtheorem{lemma}[theorem]{\textbf{Lemma}}
\newtheorem{definition}[theorem]{\textbf{Definition}}

\newcommand{\B}{\mathbb{B}}
\newcommand{\C}{\mathbb{C}}
\newcommand{\F}{\mathbb{F}}
\newcommand{\M}{\mathbb{M}}

\newcommand{\R}{\mathbb{R}}
\newcommand{\Z}{\mathbb{Z}}
\newcommand{\cC}{\mathcal{C}}
\newcommand{\cH}{\mathcal{H}}

\newcommand{\cO}{\mathcal{O}}
\newcommand{\cS}{\mathcal{S}}
\newcommand{\bA}{\mathbf{A}}
\newcommand{\bB}{\mathbf{B}}

\newcommand{\bH}{\mathbf{H}}

\newcommand{\bu}{\mathbf{u}}

\newcommand{\su}{u^{\star}}
\newcommand{\dist}{\mathrm{dist}}
\newcommand{\im}{\mathrm{Im}}
\newcommand{\rank}{\mathrm{rank}}
\newcommand{\spn}{\mathrm{Span}}
\newcommand{\supp}{\mathrm{supp}}
\newcommand{\vol}{\mathrm{vol}}

\newcommand{\qed}{\phantom{.} \hfill $\square$}

\begin{document}

\title{A Variant of the Bravyi-Terhal Bound for Arbitrary Boundary Conditions}

\author{François Arnault\thanks{XLIM, UMR 7252, Université de Limoges, France \\ Emails: \textbf{\{arnault, philippe.gaborit, nicolas.saussay\}@unilim.fr}} \and Philippe Gaborit\footnotemark[1] \and Wouter Rozendaal\thanks{IMB, UMR 5251, Université de Bordeaux, France \\ Emails: \textbf{\{wouter.rozendaal, gilles.zemor\}@math.u-bordeaux.fr}} \and Nicolas Saussay\footnotemark[1] \and Gilles Zémor\footnotemark[2] \thanks{Institut Universitaire de France}} 

\maketitle


\begin{abstract}
We present a modified version of the Bravyi-Terhal bound that applies to quantum codes defined by local parity-check constraints on a $D$-dimensional lattice quotient. Specifically, we consider a quotient $\Z^D/\Lambda$ of $\Z^D$ of cardinality $n$, where $\Lambda$ is some $D$-dimensional sublattice of $\Z^D$: we suppose that every vertex of this quotient indexes $m$ qubits of a stabilizer code $C$, which therefore has length $nm$. We prove that if all stabilizer generators act on qubits whose indices lie within a ball of radius $\rho$, then the minimum distance $d$ of the code satisfies $d \leq m\sqrt{\gamma_D}(\sqrt{D} + 4\rho)n^\frac{D-1}{D}$ whenever $n^{1/D} \geq 8\rho\sqrt{\gamma_D}$, where $\gamma_D$ is the $D$-dimensional Hermite constant. We apply this bound to derive an upper bound on the minimum distance of Abelian Two-Block Group Algebra (2BGA) codes whose parity-check matrices have the form $[\mathbf{A} \, \vert \, \mathbf{B}]$ with each submatrix representing an element of a group algebra over a finite abelian group.
\end{abstract}


\section{Introduction}

\subsection{Motivations}

In recent years, quantum Low-Density Parity-Check (qLDPC) codes have become of great interest since they are a promising ingredient for scalable quantum computing. While random matrices are commonly used to build classical LDPC codes, due to an orthogonality constraint, an algebraic ansatz is required to build quantum LDPC codes, which makes the task challenging. The last few years have seen spectacular breakthroughs, with Panteleev and Kalachev proving the existence of asymptotically good quantum LDPC codes \cite{PK22b}, see also \cite{LZ22}. However, it is not clear whether these constructions can give the best codes for short, practical lengths. For short lengths, significant efforts have gone into the search for good Two-Block codes, as is apparent from \cite{IBM24} in particular.

These quantum binary LDPC codes have parity-check matrices constructed from a pair $(\bA, \bB)$ of square commuting matrices. The first family of such codes was originally introduced by MacKay et al. \cite{MMM04}. These quantum codes are obtained by taking $\bB = \bA^\intercal$ a cyclic matrix, and were named Bicycle codes. An interesting feature of this construction is that the resulting quantum CSS codes are obtained from classical self-dual codes given by the parity-check matrices $[\bA, \bA^\intercal]$. A generalised version of this construction was presented by Pryadko and Kovalev \cite{KP13}. These Generalised Bicycle codes are obtained by taking $\bA$ and $\bB$ both cyclic. For the sake of simplicity, we will from now on also refer to these codes as Bicycles codes. A broader generalisation was presented by Pryadko and Lin \cite{LP23} in the form of Two-Block Group Algebra (2BGA) codes, constructed from a group algebra of a finite group. 

Some of these quantum codes were extensively studied in \cite{KP13, PK21, PW22} and demonstrate promising error-correcting capabilities. In the best cases, they offer almost the same error rate threshold as the surface codes, whilst improving the dimension and encoding overhead \cite{IBM24}. However, the behaviour of the minimum distance of this family of codes is unknown and remains an intriguing problem. 

Abelian Two-Block Group Algebra codes generalise Kitaev's toric codes \cite{K03, DKLP02, BD07}, since the latter correspond to the case where the abelian group is chosen to be a product of two cyclic groups. Moreover, Abelian Two-Block Group Algebra codes have the potential to yield codes with even larger minimum distances than the Kitaev code. Additionally, like toric codes, these codes can similarly be represented on lattices, albeit of higher dimensions, in such a way that the stabilizer generators of the code only act on a fixed number of qubits, all present in local regions. This feature may be appealing for physical implementations but also imply limitations on the code parameters, which are the subject of the present paper. 

\subsection{Previous results}

Due to physical constraints, a lot of attention has gone into stabilizer codes whose generators are local in a strongly geometric sense. These codes are designed so that each stabilizer generator affects only a limited number of nearby qubits. Bravyi and Terhal \cite{BT09} studied the constraints on the parameters of these codes, providing a crucial insight into the limitations of stabilizer codes. They proved that for stabilizer codes whose qubits are mapped to a region of $\Z^D$ such as $\{1, \ldots, L\}^D$, or with a periodic structure such as $(\Z / L\Z)^D$, where each stabilizer generator's support is included within a cube of $\rho^D$ vertices, then, whenever $L \geq 2(\rho-1)^2$, the minimum distance of the code cannot exceed 
\begin{equation}
\label{eq:BT}
\rho L^{D - 1} = \rho N^{\frac{D-1}{D}}
\end{equation}
where $N=L^D$ is the code length.

Subsequent research, such as \cite{BK22,BGKL23}, extended this analysis to quantum LDPC codes without explicit lattice embeddings, but with more general graph embeddings for which the subexponential behaviour of the minimum distance can also be shown, but with loosely specified constants.

Appealing to the original Bravyi-Terhal bound, Pryadko and Wang \cite{PW22} gave an upper bound on the minimum distances of Bicycle codes, a subclass of Abelian 2BGA codes. It states that the minimum distance of non-trivial Bicycle codes of length $N$ and stabilizer generators of weight $w$ are in $O(N^{\frac{D-1}{D}})$ where $D \leq w - 1$ and $D\leq w-2$ in some cases, with unspecified constants. 

\subsection{Contributions}

In the present paper, we bridge the gap between the previously known results \cite{BT09, BK22,BGKL23} on geometrically-local quantum codes and revisit the Bravyi-Terhal bound for quantum LDPC codes that can be defined by local checks on a quotient of a $D$-dimensional lattice. More precisely, consider a quantum LDPC code of length $N = nm$. If the physical qubits of the code can be indexed by the vertices of the quotient of a $D$-dimensional lattice in such a way that each vertex indexes $m$ qubits and such that all the stabilizer generators are supported within a ball of radius $\rho$, then the minimum distance of the code satisfies $d \leq m\sqrt{\gamma_D}(\sqrt{D} + 4\rho)n^\frac{D-1}{D}$ whenever $n^{1/D} \geq 8\rho\sqrt{\gamma_D}$, where $\gamma_D$ is the $D$-dimensional Hermite constant. See Theorem~\ref{thm:BT bound} for the formal statement.

Let us remark that the original Bravyi-Terhal bound \eqref{eq:BT} is stated for $m=1$ but would otherwise read as $m\rho n^\frac{D-1}{D}$. When applying general ``boundary conditions'' defined by an arbitrary lattice $\Lambda$, we therefore pay a minor penalty, that takes the form of a small constant growing at most linearly with the dimension $D$ (since $\sqrt{\gamma_D}$ is known to grow at most like $\sqrt{D}$). In the overview below, we comment on why this penalty occurs, and how it is inherent to the Bravyi-Terhal proof technique.

As an application, we provide upper bounds on the minimum distance of all Abelian 2BGA codes. Precisely, applying Theorem~\ref{thm:BT bound} with $m=2$ and $\rho=1$, we show that any Abelian 2BGA code with code length $2n$ and stabilizer generators of weight $w$, the minimum distance is bounded from above by $2\sqrt{\gamma_D}(\sqrt{D} + 4)n^\frac{D-1}{D}$ whenever $n^{1/D} \geq 8\sqrt{\gamma_D}$, where $D = w - 2$ and $\gamma_D$ is the $D$-dimensional Hermite constant. This bound is more explicit and somewhat sharper than the minimum distance bound given in \cite{PW22} for Bicycle codes: it also applies to the more general class of Abelian 2BGA codes.

\subsection{High-level overview of Theorem~\ref{thm:BT bound}}

Our proof strategy follows broadly that of Bravyi-Terhal, with some adaptations. We will partition the region $\R^D / \Lambda$, where the qubits are located, into smaller identical sets that cannot contain more than $m\sqrt{\gamma_D}(\sqrt{D} + 4\rho)n^\frac{D-1}{D}$ qubits. We will then prove that there is a non-trivial logical operator whose support is included in one of those smaller regions. The proof is divided into the following steps. 

Firstly, we identify a suitable basis for the lattice where the norm of the final vector in its associated Gram-Schmidt basis is as large as possible. This ensures we can partition $\R^D /\Lambda$ into sufficiently many regions in such a way that no stabilizer generator can simultaneously act on qubits in non-adjacent regions. These regions are translates of each other and an upper bound on their volume involves the $(D-1)$-th Rankin constant of $\Lambda$, which, by a duality argument, equals the Hermite constant $\gamma_D$. This accounts for the $\sqrt{\gamma_D}$ penalty in the modified Bravyi-Terhal bound.

Next, we repeatedly apply Bravyi and Terhal's Cleaning Lemma to demonstrate the existence of a non-trivial logical operator confined to one of these regions. As a result, the minimum distance is upper-bounded by the number of qubits contained within this region. This number is equal to $m$ times the count of integral points of that region. Depending on the given lattice, it may be hard to compute the exact number of integral points since the fundamental domain and $\Z^D$ might not be aligned appropriately. We prove that the number of integral points is bounded by the volume of an enlarged region that we can compute. This approximation costs us an additional $\sqrt D$ penalty compared to the original bound, where the number of integral points per region was just equal to its volume.

\subsection{Organisation of the paper}

The paper is divided into three parts. In Section~\ref{sec:abelian 2BGA codes}, we provide the required background on satbiliser codes and Abelian 2BGA codes. In Section~\ref{sec:BT bound}, we state and prove the variation of the Bravyi-Terhal bound for geometrically-local stabilizer codes. In Section~\ref{sec:distance bound}, we prove the aforementioned bound on the minimum distance of Abelian 2BGA codes. 


\section{Abelian Two-Block Group Algebra Codes}
\label{sec:abelian 2BGA codes}

Abelian Two-Block Group Algebra (2BGA) codes are CSS stabilizer codes constructed using the algebra over an abelian group. We briefly recall the definition of stabilizer and CSS codes and then introduce the 2BGA codes under study in this paper.

\subsection{Stabilizer and CSS codes}

The Pauli group on $N$ qubits $P_N$ is the set of all products of the form $\alpha E_1 \otimes \ldots \otimes E_N$  where $\alpha \in \lbrace \pm 1, \pm i \rbrace$ is a fourth root of unity and each $E_k$ is one of the four matrices:
\[  I = \begin{pmatrix}
1 & 0 \\ 0 & 1
\end{pmatrix}, \text{   }
X = \begin{pmatrix}
0 & 1 \\ 
1 & 0
\end{pmatrix}, \text{   }
Z = \begin{pmatrix}
1 & 0 \\
0 & - 1
\end{pmatrix}, \text{   }
Y = iXZ  = \begin{pmatrix}
0 & -i \\ i  & 0
\end{pmatrix} \in \M_2(\C).\] 
\vspace{0.1cm}

A quantum stabilizer code $C_S$ of length $N$ is a linear subspace of $(\C^2)^{\bigotimes N} $ whose elements are left invariant under the action of an abelian subgroup $S \subset P_N$ that does not contain $-I^{\bigotimes N}$: \[ C_S = \lbrace \, |\varphi \rangle \in (\C^2)^{\otimes N} \, | \, \forall E \in S, \,  E|\varphi \rangle = |\varphi \rangle \,  \rbrace.\]

It is customary to map each generator of the subgroup $S$ to a row of a binary matrix $\bH$ via the map \[\alpha \bigotimes_{i=1}^{N}X^{x_{i}}Z^{z_{i}} \mapsto (x,z) = (x_1, \ldots, x_N | z_1, \ldots, z_N).\] The resulting matrix $\bH = (\bA_X | \bA_Z)$ is called a parity-check matrix of $C$. The commutativity constraint on the generators of $ S$ translates into the equality $\bA_X \bA_Z^{\perp} + \bA_Z \bA_X^{\perp} = \mathbf{0}$.

A stabilizer code $ C_S$, whose parity-check matrix is $ \bH = (\bA_X | \bA_Z) $, has parameters $[[ N, k , d ]]$ where the matrix $\bH$ has $2N$ columns, $k = N - \rank \, \bH$ and $d$ is the smallest Hamming weight of a vector that is orthogonal to each row of $\bH$ but which does not belong to the vector space spanned by the rows of $\bH$.

\vspace{0.15cm}
A quantum Calderbank-Shor-Steane (CSS) code \cite{CS96, S96} is a stabilizer code for which each generator is a product of only bit-flip operators $X$ or a product of only phase-flip operators $Z$. It follows that its parity-check matrix is of the form \[\bH = \begin{pmatrix}
\bH_X  & \mathbf{0} \\ 
\mathbf{0} & \bH_Z
\end{pmatrix}\]
where the matrices $ \bH_X$ and $ \bH_Z$ satisfy the condition $\bH_X  \bH_Z^\intercal = \mathbf{0}$.

The length $N$ of a CSS code, or the number of physical qubits, is the number of columns of $\bH_X$ (or of $\bH_Z$). Its dimension $k$, or the number of logical qubits, is $k = N - \rank \, \bH_X - \rank \, \bH_Z$. The minimum distance $d$ of the code is defined as $\min(d_X,d_Z)$, where $d_X$ (resp. $d_Z$) is equal to the smallest Hamming weight of a non-zero vector of $\ker \bH_X$ (resp. $\ker \bH_Z)$ that is not in the row-space of $\bH_Z$ (resp. $\bH_X)$. 

\subsection{Abelian 2BGA codes}
\label{sec: 2BGA codes}

Inspired by the previously known Bicycles codes \cite{KP13}, Abelian Two-Block Group Algebra (2BGA) codes were proposed by Priadko and Lin \cite{LP23}. These quantum binary LDPC codes are defined by parity-check matrices constructed from a pair of square commuting matrices obtained from a group algebra of a finite abelian group.

Let $G = \{ g_1, \dots, g_n \} $ be a finite abelian group of order $n$ and consider the group algebra $\F_2[G]$ of all formal linear combinations of elements of $G$ with coefficients in $\F_2 $. For $g \in G$, consider the the permutation matrix $\B(g) \in \M_n(\F_2)$, defined by  $\B(g)_{i,j} = 1$ if and only if $g_i = gg_j$, and which describes the action of $g$ on $G$. The product of two such matrices is commutative: $\B(gh) = \B(g)\B(h)$ for all $g,h \in G$. 

An Abelian 2BGA code is a CSS stabilizer code defined by two binary matrices $\bH_X = [\bA | \bB]$, $\bH_Z = [\bB^{\intercal} | \bA^{\intercal}]$ where $\bA = \sum_{g \in G} a_g \B(g)$ and $\bB = \sum_{g \in G} b_g \B(g)$, with $ a_g, b_g \in \F_2$. Such a code is non-trivial whenever $\bA$ or $\bB$ is non-zero. The code length is $N = 2n$. Clearly $\rank \, \bH_X = \rank \, \bH_Z$ and so the code dimension $k$ is even \cite{PK20}. Moreover, we have $d_X = d_Z$ since the Abelian 2BGA codes defined by the matrices $(\bA,\bB)$ and $(\bB,\bA)$ are equivalent \cite{LP23}.

When the abelian group is taken to be cyclic, one obtains the Bicycle codes introduced by Pryadko and Kovalev in \cite{KP13}. If furthermore $\bA=\bB$, then one recovers the Bicycle Codes of MacKay et al. \cite{MMM04}. When the abelian group is the product of two cyclic groups, i.e. of the form $\Z_r \times \Z_s$, then one obtains the Bivariate Bicycle codes \cite{IBM24}. Finally, note that Abelian 2BGA codes are the simplest form of Lifted Product codes introduced by Panteleev and Kalachev \cite{PK22a}, with both matrices of dimension $1 \times 1$. 


\section{Bravyi-Terhal bound}
\label{sec:BT bound}

Before stating and proving our bound on the minimum distance of geometrically-local stabilizer codes, we recall some background on lattices.

\subsection{Lattices}

\begin{definition}[Lattices] 
A lattice of $\R^D$ is discrete additive subgroup $\Lambda$ of $\R^D$ of the form \mbox{$\Lambda = \sum_{i=1}^m \, \Z \bu_i$} where $\bu_1, \dots, \bu_r \in \R^D$ are $\R$-linearly independent vectors. The family $\{\bu_1, \dots, \bu_r\} $ is called a basis of $\Lambda$, and the integer $r$ is called the rank of the lattice. When $r = D$, we say that the lattice has full rank. The determinant of a lattice $\Lambda$ is $\det(\Lambda) = \sqrt{\det(\bB^{\intercal}\bB)}$ for any basis $\bB$ of $\Lambda$, and corresponds to the volume of $\mathcal{P}(\bu_1, \dots, \bu_r) = \{\sum_{i=1}^r x_i \bu_i \, \vert \, 0 \leq x_i < 1\}$, the fundamental domain of $\Lambda$ with respect to $\bB$. Finally, when $\Lambda$ has full rank, $\det(\Lambda) = |\det(\bB)|$.
\end{definition}

In addition to their geometrical properties, we use the topological properties of lattices. Indeed, any full rank lattice defines a regular tiling of $\R^D$ where each tile is obtained by translating the fundamental domain of $\Lambda$ by an element of $\Lambda$, i.e. $\R^D = \coprod_{g \in \Lambda} \, (g + \mathcal{P})$. Since lattices are also discrete subgroups of $\R^D$, the quotient of $\R^D$ by any full-rank lattice is a metric space. 

\begin{proposition}
Let $\Lambda \subset \R^D$  be a full rank lattice and $ \| \cdot \| $ be any norm on $ \R^D $. Then $ \R^D/\Lambda $ is a metric space for the distance $ dist $ induced by $ \| \cdot \|$:
\begin{equation}
\begin{array}{cccc}
dist    : &  \R^D / \Lambda   \times   \R^D / \Lambda  & \longrightarrow  & \R^+ \\
 &   ( [a], [b] )  & \mapsto &  \,  inf_{g \in \Lambda}  \, \,   \| a - (b+g) \|
\end{array}
\end{equation}
Moreover, for any pair $(a,b) \in \R^D \times \R^D$ there exists $ g_0 \in \Lambda $ such that $ dist([a], [b]) = \| a - (b + g_0) \|$.
\end{proposition}

\vspace{0.15cm}
If we have a lattice $\Lambda \subseteq \Z^D$  whose volume is large enough and a stabilizer code whose qubits can be represented on the quotient of $ \Z^D/ \Lambda $ in such a way that each generator of the code only acts locally on neighbouring qubits, then we prove that the minimum distance of the code is bounded by $\kappa_D \vol(\Lambda)^{\frac{D-1}{D}}$ where $\kappa_D $ is a constant that depends on the locality of the generators, on the dimension $D$, and on the $D$-dimensional Hermite constant that we recall below.

\subsection{Hermite and Rankin constants}

\begin{definition}[Hermite constant \cite{H1850}]
For a full rank lattice $\Lambda \subset \R^D$, consider \[\gamma_{D}(\Lambda) = \min\{\left( \frac{\|x\|)}{\vol(\Lambda)^\frac{1}{D}} \right)^2 \, \vert \, x \in \Lambda \text{ with } x \neq 0\}.\] The $D-$dimensional  Hermite constant is defined by $\gamma_{D} = \max\{\gamma_{D}(\Lambda) \, \vert \, \Lambda \text{ rank $D$ lattice}\}$.
\end{definition}

The first linear upper bound on Hermite’s constant is a direct consequence of Minkowski's Convex Body Theorem. 

\begin{theorem}[Minkowski]
\label{thm: Minkowski}
Let  $\Lambda$  be a full-rank latice of $\R^D$ and let $S \subseteq \R^D$ be a symetric, convex body such that $\vol(S) > 2^D \vol(\Lambda)$. Then there is a non-zero element of $ \Lambda $ in $S$. 
\end{theorem}

Minkowski's theorem shows that any ball whose volume is sufficiently large contains at least a nonzero lattice vector. This gives directly an upper bound on the shortest length of a nonzero lattice vector and from this bound it is possible to derive a linear bound on the Hermite constant: \[ \gamma_D \leq 1 + \frac{D}{4}. \]

\vspace{0.15cm}
Rankin \cite{R56} provided a generalisation of Hermite constants. 

\begin{definition}[Rankin constant]
Let $ \Lambda \subset \R^D$ be a full-rank lattice and let $1 \leq r \leq D$,
\begin{equation}
\label{eq: constante de Rankin d'un réseau L}
\gamma_{D,r}(\Lambda) = \min\{\left( \frac{\vol(\{x_1, \ldots, x_r \})}{\vol(\Lambda)^\frac{r}{D}} \right)^2 \, \vert \, \{x_1,\ldots,x_r \} \in \Lambda \text{ with } \vol(\{x_1,\ldots,x_r \}) \neq 0\}.
\end{equation}
\end{definition}
The Rankin constants are defined by $\gamma_{D,r} = \max\{\gamma_{D,r}(\Lambda) \, \vert \, \Lambda \text{ rank $D$ lattice}\}$. By a duality argument \cite{M02}, it can be shown that $\gamma_{D,r} = \gamma_{D,D-r}$. Moreover, note that $\gamma_{D,1} $ is clearly equal to $ \gamma_{D}$, the $D$-dimensional Hermite constant.

\subsection{Geometrically-local codes}

A stabilizer code is said to be geometrically-local if one can index its physical qubits by the vertices of a lattice in such a way that the code's stabilizer generators act only on neighbouring qubits. In \cite{BT09}, Bravyi and Terhal showed that when such a geometrically-local code can be mapped to a region of $\Z^D$ such as $\{1, \ldots, L\}^D$, or with a periodic structure such as $(\Z / L\Z)^D$, an upper bound on the minimum distance of the code is given by \eqref{eq:BT}. We consider a variation of their theorem, for which the code is mapped on a quotient $\Z^D / \Lambda$, where $\Lambda$ is an arbitrary $D$-dimensional sublattice of $\Z^D$.

\begin{theorem}
\label{thm:BT bound}
Let $\cC$ be stabilizer code of length $N = mn$. Suppose that the following conditions are satisfied:
\begin{itemize}
\item The qubits are indexed by the vertices of $\Z^D / \Lambda$, where $\Lambda$ is a $D$-dimensional sublattice of $\Z^D$ such that $\vert \Z^D / \Lambda \vert = n$, and each vertex indexes $m$ qubits.
\item The parity-checks are geometrically-local: there exists $\rho > 0$ such that the support of any stabilizer is contained in a Euclidean ball of $\Z^D / \Lambda $ of radius $\rho$.
\item $n^{1/D} \geq 8\rho \sqrt{\gamma_D}$ where $ \gamma_D $ is the $D$-dimensional Hermite constant.
\end{itemize}
Then the minimum distance of the stabilizer code $\cC$ satisfies $d < m  \sqrt{\gamma_D} (\sqrt{D} + 4\rho) n^\frac{D-1}{D}$.
\end{theorem}

\subsection{Proof of Theorem~\ref{thm:BT bound}}

Let us start by recalling the Cleaning lemma that was first introduced by Bravyi and Terhal in \cite{BT09}. Let $\cC$ be a stabilizer code with stabilizer group $S$. Let $M$ denote a subset of the $N$ qubits in the code block. If $\cO$ is one of the code’s logical operators, we say that $\cO$ can be cleaned on $M$ if there is a logically equivalent operator $\tilde{\cO} = \cO \cS$, where $\cS$ is an element of the code stabilizer $S$, such that $\tilde{\cO}$ acts trivially on $M$. 

\vspace{0.15cm}
\begin{lemma}(Cleaning Lemma)
\label{lem:cleaning lemma}
If there is no non-trivial logical operator whose support can be contained in $M$, then any logical operator can be cleaned on $M$.
\end{lemma}

\vspace{0.15cm}
Let $\cC$ be a stabilizer code of length $N = mn$ and let $\Lambda$ be a $D$-dimensional sublattice of $\Z^D$ such that $\vert \Z^D / \Lambda \vert = n$, each vertex indexes $m$ qubits, and there exists $\rho > 0$ such that the support of any stabilizer generator is contained in a Euclidean ball of $\Z^D / \Lambda $ of radius $\rho$.

\vspace{0.15cm}
To prove Theorem~\ref{thm:BT bound}, we shall use the Cleaning Lemma repeatedly. We therefore need to define appropriate regions that partition the qubits of the code $\cC$. The first step of the proof is to choose an appropriate basis for our lattice. In our case, it is a basis whose associated Gram-Schmidt basis has a vector of large enough length. A convenient way to find such basis is to use the work of \cite{N09} in order to build a family $\{ u_1, \dots, u_{D-1} \} $ of linearly independent vectors of $ \Lambda$ that can be completed into a basis of $ \Lambda $ and which is such that the volume of the $D - 1$ dimensional lattice generated by $ u_1, \dots, u_{D-1}$ is minimal.

\vspace{0.15cm}
\begin{lemma}
\label{lemma: good basis for Lambda}
There exists a family of linearly independent vectors $u_1, \ldots, u_{D-1} \in  \Lambda$ that can be extended into a basis of $\Lambda$ and satisfies:
\begin{equation}
\label{eq:rankin}
0 < \left( \frac{\vol(\{u_1, \ldots, u_{D-1} \})}{n^\frac{D-1}{D}} \right)^2 = \gamma_{D,D-1}(\Lambda) \leq \gamma_D
\end{equation}

where
\begin{itemize}
\item $ \vol(\{u_1, \ldots, u_{D-1} \}) $ is the volume of the lattice generated by $ u_1, \dots, u_{D-1}$,
\item $\gamma_{D,D-1}(\Lambda)$ is the minimum of $\vol(S)^2 \, n^{-2 \, \frac{D-1}{D}} $ over all rank $(D-1)$ sublattice $S$ of $ \Lambda $,
\item $ \gamma_D $ is the $D$-dimensional Hermite constant. 
\end{itemize}
\end{lemma}
\vspace{0.15cm}

\vspace{0.15cm}
\textit{Proof:} From the definition of the Rankin constant \cite{M02}, we have that $ \gamma_{D,D-1}(\Lambda) \leq \gamma_D $ and that there is a rank $(D-1)$ sublattice  $S_0 \subset \Lambda $ such that $ 0 < \gamma_{D, D-1}(\Lambda) = \vol(S)^2 \, n^{-2 \, \frac{D-1}{D}} $. We now need to find a basis $ u_1, \dots, u_D $ of $ \Lambda $ such that $ \vol(S_0) = \vol( \{ u_1, \dots, u_{D-1} \} )$. We prove the existence of this basis by using the structure theorem for finitely generated modules over the principal ideal domain $\Z$. Since $\Lambda $ and $S_0 $ are respectively lattices of rank $D$ and $D - 1$, there exist a basis  $ \{ u_1, \dots, u_D \}$ of $ \Lambda$ and integers $ x_1, \dots, x_{D-1} \in \Z \backslash \{0\} $ such that $ \{ x_1u_1, \dots, x_{D-1}u_{D-1}  \} $ is a basis of $ S_0$. We have that the volume of $ S_0$ is equal to  $ \vol(\{u_1, \dots, u_{D - 1} \}) $. Indeed, let $T$ be the $D - 1$ lattice generated by $\{u_1, \dots, u_{D-1}\}$. On one hand, by minimality of the volume of $S_0$, we have that $ \vol(S_0) \leq \vol(T) $. On the other hand, since $ S_0 $ is generated by $\{ x_1u_1, \dots, x_{D-1}u_{D-1} \}$, we have $\vol(S_0) = |T/S_0| \vol(T) \geq \vol(T)$. It therefore follows that $\vol(S_0) = \vol(T) = \vol( \{ u_1, \dots, u_{D-1} \} )$. \qed

\vspace{0.15cm}
Let $ u_1, \dots, u_D \in \Lambda $ be a basis of $ \Lambda $ such that $\vol(u_1, \dots, u_{D-1}) = \sqrt{\gamma_{D, D-1}(\Lambda)} n^\frac{D-1}{D}$. This particular basis has the property that the length of the last vector of the associated orthogonal Gram-Schmidt basis $\{\su_1, \ldots, \su_D\}$ is as large as possible.

\vspace{0.15cm}
\begin{lemma}
\label{lem:bound on the last gram-schmidt vector}
$\|\su_D\| \geq \frac{n^{1/D}}{\sqrt{\gamma_D}} $.
\end{lemma}

\vspace{0.15cm}
\textit{Proof:} The lattice $\Lambda$ has volume $n = \prod_{i=1}^{D} \|\su_i\|$. By considering the sublattice generated by the first $D-1$ basis vectors, we have $n = \vol(\{\su_1, \ldots, \su_{D-1}\})\|\su_D\| = \vol(\{u_1, \ldots, u_{D-1}\})\|\su_D\|$ and so $\|\su_D\| = \frac{n}{\vol(\{u_1, \ldots, u_{D-1}\})}$. Since $ \vol(\{u_1, \ldots, u_{D-1}\}) = \sqrt{\gamma_{D, D-1}(\Lambda)} n^\frac{D-1}{D}$ and that \mbox{$0 < \gamma_{D, D-1}(\Lambda) \leq \gamma_D$}, we have that $\|\su_D\| \geq \frac{n^{1/D}}{\sqrt{\gamma_D}}$. \qed

\vspace{0.15cm}
We now partition $\R^D / \Lambda$ into $\mu$ parallelotopes $(T_k/\Lambda)_{0 \leq k < \mu}$, where \[T_k = \{x =  \sum_{i=1}^{D} x_i u_i \in \R^D  \, \vert \,\forall i \in [[ 1, D - 1 ]], \,  x_i \in [0 ; 1) \text{  and  } \, x_D \in [\frac{k}{\mu} ; \frac{k+1}{\mu}) \, \}.\]
and $\mu$ is an even integer whose exact value we will determine later.

These parallelotopes are obtained by dividing the fundamental domain $\mathcal{P}$ of the lattice $\Lambda$ into $\mu$ regions delimited by $\mu + 1$ hyperplanes $\cH_k$, defined by translating the hyperplane \mbox{$\cH_0 = \spn \{u_1, \ldots, u_{D-1}\}$} by the vectors $\frac{k}{\mu} u_D$ for $k \in [[0; \mu]]$.
The \textit{width} $\lambda$ of these parallelotopes is then defined to be the distance separating two consecutive hyperplanes. Since $\su_D$ is orthogonal to $\cH_0$, the width of a parallelotope is thus given by $\lambda = \frac{ \| \su_D \| }{ \mu }$.

\vspace{0.15cm}
If this partition has an even number of parallelotopes and  is such that no stabilizer generator of the code acts on more than two adjacent parallelotopes, then the Cleaning Lemma provides an upper bound on the minimum distance of our geometrically-local stabilizer code $\cC$.

\vspace{0.15cm}
\begin{lemma}
\label{lem:bound on d}
If the partition contains an even number of parallelotopes and if no stabilizer generator of the code acts on more than two adjacent parallelotopes, then there exists a non-trivial logical operator whose support is contained in one of the parallelotopes. Hence, the minimum distance $d$ of the code satisfies $d \leq m|(\Z^D \cap T_k) / \Lambda|$.
\end{lemma}
\vspace{0.15cm}
\textit{Proof:} Let us assume that we have partition $ \R^D/\Lambda$ into an  even number $ \mu $ of parallelotopes $ T_k/\Lambda$ and that no stabilizer generator of the code acts on more than two adjacent parallelotopes.

If there is a non-trivial logical operator whose support is included in one of the odd index parallelotope $ T_{2k + 1}/ \Lambda $ then the minimum distance is upper bounded by $ d \leq m |(Z^D \cap T_{2k + 1}) / \Lambda |$. Otherwise, we consider a non-trivial logical operator $\cO$. By the cleaning lemma, this operator may be cleaned out on each odd index parallelotope $ T_{2k+1}/\Lambda$ individually. Moreover, since no stabilizer generator of the code acts on more than two adjacent parallelotopes, none of them can act on two parallelotopes with odd indices  at the same time. Hence, we can use the Cleaning Lemma multiple times in order to clean out $ \cO $ on each of these odd index parallelotopes. By doing so, we are left with an equivalent non-trivial logical operator $\tilde{\cO}$ whose support is included in the reunion of the parallelotopes with even indices. We can write, $ \tilde{\cO} = \tilde{\cO}_{T_0} \times \tilde{\cO}_{T_2} \times \dots \times \tilde{\cO}_{T_{\mu - 2}} $ where each $ \tilde{\cO}_{T_{2i}} $ is an element of the Pauli group $ \mathcal{P}_n$ whose support is included in  $T_{2i}/\Lambda $. If $ \mu = 2$ then  $\tilde{\cO} = \tilde{\cO}_{T_0} $ is a non-trivial logical operator whose support is included in $T_0$. Thus, $d \leq m|(\Z^D \cap T_0)/\Lambda|$. Otherwise, we set $p \in [[ 0, \frac{\mu - 2}{2} ]]$ and note that each stabilizer generator $\cS \in S$ commutes with $ \tilde{\cO}_{T_{2p}}$. Indeed, for an arbitrary stabilizer generator $\cS$, either $\cS$ does not act on any qubit of $T_{2p}/\Lambda$ and so $\cS$ commutes with $\tilde{\cO}_{T_{2p}}$, or either $\cS$ acts on some qubits of $T_{2p}/\Lambda$ but then $\cS$ commutes with $ \tilde{\cO}_{T_{2i}}$ for $i \in [[ 0, \frac{\mu - 2}{2} ]] \backslash \{p\}$ because no stabilizer generator of the code acts on more than one parallelotope with even index. Since, $ \tilde{\cO} = \prod_{i = 0}^{\frac{\mu - 2}{2}} \tilde{\cO}_{T_{2i}}$ is logical, it commutes with the stabilizer generator $\cS$ and thus, $\tilde{\cO}_{T_{2p}}$ also has to commute with $\cS$. 

Finally, as $ \tilde{\cO}$ is a product of $ \tilde{\cO}_{T_{2q}}$ and is not a stabilizer, one of its factor must also not be a stabilizer. Thus, there is a factor $ \tilde{\cO}_{T_{2q}} $ which is a non-trivial logical operator and whose support is contained in $T_{2q}/\Lambda $. Therefore, $d \leq m|(\Z^D \cap T_{2q})/\Lambda|$. \qed

\vspace{0.15cm}
Before explaining how we choose the constants $\mu$ and $\lambda$ to apply Lemma~\ref{lem:bound on d}, we give an upper bound on the number of integral points in the parallelotope $T_k/\Lambda$ from which the bound on the minimum distance of Theorem~\ref{thm:BT bound} follows. 

\vspace{0.15cm}
\begin{lemma}
\label{lem:bound on integral points}
$|(\Z^D \cap T_k)/\Lambda | \leq \frac{n}{\|\su_D\|}(\lambda + \sqrt{D})$.
\end{lemma}

\textit{Proof:} To bound the number of integral points in the parallelotope $T_k/\Lambda$, we first prove that the number of integral points is equal to the volume of a set $ C/\Lambda$, corresponding to the intersection of the fundamental domain  $\mathcal{P}$ with the reunion of $\cup_{g \in \Lambda} g + C $. Secondly, we show that $C /\Lambda $ is included in a set $ \Gamma/\Lambda$ whose volume does not exceed $\frac{n}{\|\su_D\|}(\lambda + \sqrt{D})$. 
\vspace{0.15cm}

Noting that the $D$-dimensional unit cubes $C_x = \{ t \in \R^D \, \vert \, \| t - x \|_\infty < \frac{1}{2}  \}$ centered at each point $x \in \Z^D \cap T_k$  are pairwise disjoint and that the volume of each $ C_x $ modulo $ \Lambda $ is equal to $1$, it follows that: \[ |\Z^D \cap T_k | = \sum_{x \in \Z^D \cap T_k} \vol(C_x/\Lambda) = \vol \left( \cup_{x \in \Z^D \cap T_k} \,  (C_x/\Lambda) \,  \right). \] Depending on the boundary conditions, the union of all the cubes $\bigcup_{x \in \Z^D \cap T_k} C_x/\Lambda $ may not be included in the parallelotope $T_k/\Lambda$. Indeed, if one considers an integral point nearby the border of the parallelotope, then the corresponding cube may not entirely lie inside the parallelotope. However, since the diagonal of a $D$-dimensional unit cube is $\sqrt{D}$, we always have that $ \cup_{x \in \Z^D \cap T_k} (C_x/\Lambda) \subseteq \Gamma / \Lambda $ where $\Gamma$ is defined by \[\Gamma = \{\,  \sum_{i = 1}^D \, a_iu_i \, \, | \, \,  \forall i \in [[ 1, D - 1 ]], 0 \leq a_i < 1 \, \text{  and  } \, \frac{-\sqrt{D}}{2\|u_D^*\|} + \frac{k}{\mu}    \leq a_D < \frac{\sqrt{D}}{2\|u_D^*\|} + \frac{k+1}{\mu}\}. \] Thus $ |\Z^D \cap T_k | \leq \vol(\Gamma/ \Lambda)$. Next, by using the fact that the $D$-dimensionnal volume is sub-additive, translation invariant and that we have a regular tiling $\R^D = \coprod_{g \in \Lambda} g + \mathcal{P} $, we show that the volume of $ \Gamma/ \Lambda $ is less than the $ D$-dimensionnal volume of $\Gamma $:
\[\vol(\Gamma/ \Lambda)  = \vol(\, \mathcal{P} \, \cap \,  \cup_{g \in \Lambda} (g + \Gamma)  \,  ) \leq \sum_{g \in \Lambda} \vol( \mathcal{P}  \cap \Gamma  + g ) = \vol(\Gamma).\]

The set $\Gamma$ is obtained by extending the parallelotope $T_k$ by $\frac{\sqrt{D}}{2\| u_D^* \|} $ on both side given by the direction $u_D^*$. Hence we have that
\[ |\Z^D \cap T_k|  \leq \vol(\Gamma) =   n ( \frac{\sqrt{D}}{\| u_D^* \| } + \frac{1}{\mu} ) = \frac{n}{\| u_D^*\|}( \sqrt{D} + \lambda) \text{ because $ \| \su_D \| = \lambda \mu $.} \]

Finally, there is a one-to-one correspondance between $ \mathcal{P} $ and $ \R^D/\Lambda$, so considering that $T_k$ is a subset of $ \mathcal{P}$, the number of integral points of $ T_k $ inside or outside of the quotient are the same, that is $|(\Z^D \cap T_k) / \Lambda | = |\Z^D \cap T_k |$. \qed
\vspace{0.15cm}

We now explain how to choose $\lambda $ and $\mu$. Recall that there are two requirements: we need an even number $\mu$ of parallelotopes to apply the Cleaning Lemma and the width $\lambda $ of each parallelotope has to be sufficiently large to ensure that each stabilizer generator acts only on at most two adjacent parallelotopes. The next proposition give widths for which the condition on the stabilizer generators support is always satisfied.

\vspace{0.15cm}
\begin{lemma}
\label{lem:BT condition}
If $\lambda \geq 2\rho$, then the support of any stabilizer generator is included in at most two adjacent parallelotopes.
\end{lemma}

\vspace{0.15cm}
\textit{Proof:}
The distance in the quotient $\Z^D / \Lambda$ between two distincts parallelotopes $T_i/\Lambda $ and $T_j/\Lambda $ is given by the minimum of the distances separating the hyperplanes delimiting both parallelotopes. Hence it corresponds to the minimum of the distances in the quotient $\Z^D / \Lambda$ between $\cH_{i+1}$ and $\cH_{j}$, or between $\cH_{i}$ and $\cH_{j+1}$. Let $|.|_\mu$ denote the absolute value modulo $\mu$, i.e. $|x|_\mu$ corresponds to the smallest non-negative integer who is congruent to $x$ modulo $\mu$. Then for $i \neq j$, we have $\dist(T_i/\Lambda,T_j/\Lambda) = \min\{|i-j-1|_\mu,|j-i-1|_\mu\} \lambda$. The support $S$ of a stabilizer generator is included in a ball of radius $\rho$, hence the distance separating any two qubits of $S$ is less than $2\rho$. If $S$ is included in strictly more than two adjacent parallelotopes, then there exist  $i$ and $j$ such that   $|i-j|_\mu > 1$  and $[q_1] \in S \cap T_i$ and $[q_2] \in S \cap T_j$. Then $\dist([q_1],[q_2]) > \dist(T_i/\Lambda ,T_j/\Lambda) \geq \lambda \geq 2\rho$  which contradicts our assumption on $ \lambda$. \qed

\vspace{0.15cm}
To apply Lemma~\ref{lem:bound on d}, the requirements are that the thickness $\lambda$ of the parallelotopes should be at least of $2\rho$ and that the number of parallelotopes $\mu$ should be even.  Let $\ell = \| \su_D \| $ denote the norm of $\su_D$. We may chose $\lambda$ and $\mu$ as follows: \[\lambda = \frac{\ell}{\mu} \text{ with } \mu = \left\{\begin{array}{ll}
\lfloor \frac{\ell}{2\rho} \rfloor & \text{if it is even}, \\
\lfloor \frac{\ell}{2\rho} \rfloor - 1 & \text{otherwise}.
\end{array} \right.\] 

As desired we have $\lambda \geq 2\rho$ since $\mu \leq \lfloor \frac{\ell}{2\rho} \rfloor \leq \frac{\ell}{2\rho}$, and provided that $ \ell $ is large enough, $\mu $ is even.

The smaller the parallelotope width $\lambda $ is, the better the bound on the minimum distance of Theorem \ref{thm:BT bound} is. Thus,  we would like $\lambda$ to be as small as possible and hence $\mu$ to be as large as possible. The following Lemma asserts that when the volume of the lattice is large enough, $ \lambda$ is upper bounded  by $ 4\rho $.

\vspace{0.15cm}
\begin{lemma}
\label{lem:bound on lambda}
If $n^{1/D} \geq 8\rho\sqrt{\gamma_D}$, then $ \mu \geq 2$ and $\lambda < 4\rho$.
\end{lemma}

\vspace{0.15cm}
\textit{Proof:} If $n^{1/D} \geq 8\rho\sqrt{\gamma_D}$, then from Lemma~\ref{lem:bound on the last gram-schmidt vector}, we have $ \ell \geq \frac{n^{1/D}}{\sqrt{\gamma_D}} \geq 8 \rho $. Hence we obtain $\mu \geq \lfloor \frac{\ell}{2\rho} \rfloor - 1 >  \frac{\ell}{2\rho} - 2 \geq 2$. It follows that $\lambda = \frac{\ell}{\mu } < \frac{2\rho\ell}{\ell - 4\rho} \leq 4 \rho $ since $ 0 < 2 \rho \ell \leq 4 \rho ( \ell - 4 \rho ) $. \qed

\vspace{0.15cm}
We have now all the ingredients to prove Theorem~\ref{thm:BT bound}.

\vspace{0.15cm}
\textit{Proof of Theorem~\ref{thm:BT bound}}:
Let $\cC$ be a stabilizer code of length $mn$ such that there exists $\Lambda$ a $D$-dimensional sublattice of $\Z^D$ such that $\vert \Z^D / \Lambda \vert = n $ and that each vertex of $ \Z^D/\Lambda $ indexes $m$ qubits. Assume that there exists $\rho > 0$ such that the support of any stabilizer generator is contained in a Euclidean ball of $\R^D / \Lambda $ of radius $\rho$ and that $ n^\frac{1}{D} \geq 8 \rho \sqrt{\gamma_D} $ where $ \gamma_D $ is the Hermite constant. Let $d$ be the minimum distance of the code.

We have shown that we can partition $\R^D / \Lambda $ into an even number $\mu$ of parallelotopes $T_k/\Lambda$ of width $\lambda$ that satisfies $2\rho \leq \lambda < 4 \rho $. By Lemma~\ref{lem:BT condition}, the support of any stabilizer generator is included in at most two adjacent parallelotopes. Hence, by Lemma~\ref{lem:bound on d}, there exists \mbox{$k \in [[0;\mu-1]]$}, such that $d \leq m|(\Z^D \cap T_k)/\Lambda|$. Thus according to Lemma~\ref{lem:bound on integral points}, $d \leq m\frac{n}{\|\su_D\|}(\lambda + \sqrt{D})$. By Lemma~\ref{lem:bound on the last gram-schmidt vector}, $\|\su_D\| \geq \frac{n^{1/D}}{\sqrt{\gamma_D}}$ so $d \leq m\sqrt{\gamma_D} (\lambda + \sqrt{D})n^\frac{D-1}{D}$. Finally, by Lemma~\ref{lem:bound on lambda}, we get $\lambda < 4\rho$ and thus $d < m(\sqrt{D} + 4\rho) \sqrt{\gamma_D} \, n^\frac{D-1}{D}$. \qed

\section{Minimum distance bound for Abelian 2BGA codes}
\label{sec:distance bound}

We now want to apply the previous upper bound on the minimum distance of geometrically-local codes to the case of non-trivial Abelian 2BGA codes. By showing that any Abelian 2BGA code is a geometrically-local code satisfying the conditions of Theorem~\ref{thm:BT bound}, we obtain the following result:

\begin{theorem}
\label{thm:quantum upper bound}
The minimum distance of a non-trivial Abelian 2BGA code of length $n$ and stabilizer generators of weight $ w$, is bounded from above by $2 \sqrt{\gamma_D}(\sqrt{D} + 4)n^\frac{D-1}{D}$ whenever $n^{1/D} \geq 8\sqrt{\gamma_D}$, where $D = w-2$ and $\gamma_D$ is the Hermite constant.
\end{theorem}

\vspace{0.15cm}
To prove Theorem~\ref{thm:quantum upper bound}, we first show that any non-trivial 2BGA code is equivalent to a 2BGA code generated by two binary square matrices $\bA$ and $\bB $ of size $n$, having a non-zero entry at a given position. Then, we show that the physical qubits of this latter 2BGA code can be mapped by pairs on the vertices of a quotient $\Z^D / \Lambda$, in such a way that the stabilizer generators of the code are geometrically-local of radius $1$. We then apply Theorem~\ref{thm:BT bound} with $m=2$ and $\rho=1$ to conclude. 

\vspace{0.15cm}
\begin{lemma} 
Any non-trivial Abelian 2BGA code is equivalent to an Abelian 2BGA code whose parity-chek matrix is of the form $ \bH_X =  [\bA | \bB]$ and $\bH_Z = [\bB^{\intercal} | \bA^{\intercal}]$ where $\bA = \sum_{g \in G} a_g \B(g)$ and $\bB = \sum_{g \in G} b_g \B(g)$ with $ a_e = b_e = 1 $.
\end{lemma}

\vspace{0.15cm}
\textit{Proof:}
Let $\cC$ be a non-trivial Abelian 2BGA code defined by the matrices $\bA = \sum_{g \in G} a_g \B(g)$ and $\bB = \sum_{g \in G} b_g \B(g)$ associated to the non-zero elements $ a = \sum_{g \in G} a_g g $ and $ b = \sum_{g \in G} b_g g $ of the algebra group $ \F_2[G]$. Consider $g_a, g_b \in G$ s.t. $a_{g_a} = b_{g_b} = 1$. We now consider the elements $\tilde{a} = g_a^{-1} a = e  + \sum_{g \in G \setminus \{g_a\}} a_g g$ and $\tilde{b} = g_b^{-1} b = e  + \sum_{g \in G \setminus \{g_b\}} b_g g$. The matrices constructed from $\tilde{a}$ and $\tilde{b}$ are respectively the matrices $\tilde{\bA} = \sum_{g \in G} a_g \B(g_a^{-1} g) = \B(g_a^{-1}) \sum_{g \in G} a_g \B(g) = \B(g_a^{-1}) \bA$ and $\tilde{\bB} = \B(g_b^{-1}) \bB$. The endomorphism of $\F_2^{2n}$ given by $(u \hspace{0.2cm} v) \mapsto (\B(g_a^{-1})u \hspace{0.2cm} \B(g_b^{-1})v)$ is an ismorphism that preserves the Hamming metric and induces bijections between the kernels of $[\bA, \, \bB]$ and $[\B(g_a^{-1}) \bA, \, \B(g_b^{-1}) \bB]$, and between the row-spaces of $[\bB^\intercal, \, \bA^\intercal] $ and $[(\B(g_b^{-1}) \bB)^\intercal, \, (\B(g_a^{-1}) \bA)^\intercal]$. Hence the Abelian 2BGA codes defined by $(\bA, \bB)$ and $(\tilde{\bA}, \tilde{\bB})$ are equivalent. Consequently, to study the minimum distance of the code, we may assume that $e \in \supp(a)$ and $e \in \supp(b)$. \qed

\vspace{0.15cm}
We may now suppose that $\supp(a) = \{g \in G \, | \, a_g \neq 0 \}= \{e, g_{a_1}, \ldots, g_{a_r}\}$ and similarly that $\supp(b) = \{e, g_{b_1}, \ldots, g_{b_s}\}$. Consider now the $\Z$-linear map \[\Psi : \Z^r \times \Z^s \rightarrow G, \; \epsilon_i \mapsto \left\{\begin{array}{ll}
g_{a_i} & \text{if } 1 \leq i \leq r, \\
g_{b_{i-r}} & \text{if } r+1 \leq i \leq r+s,
\end{array} \right.\] where $\{\epsilon_i\}_{1 \leq i \leq r+s}$ is the computational basis of $\Z^{r+s}$.

\vspace{0.15cm}
\begin{lemma}
If the Abelian 2BGA code $\cC$ cannot be decomposed into a direct sum of smaller  Abelian 2BGA codes, then $\Psi$ is surjective. It therefore induces an isomorphism $\Bar{\Psi}$ between $ \Z^{r+s}/ \ker \Psi$ and $G$.
\end{lemma}

\vspace{0.15cm}
\textit{Proof:}
Consider the subgroup given by the image of $\Psi$, $H := \im(\Psi) \, \subset G$. If $H \neq G$, then the code is a copy of $[G:H]$ codes of length $2|H|$. Indeed, one can reorder the elements of $G$ in such a way that they represent the successive cosets of $G/H$. Since $a,b \in \F_2[H]$, $\forall g \in G$, $ga, gb \in \F_2[gH]$. Hence $\bA$ and $\bB$ are block diagonal matrices and so by considering each pair of blocks of $\bA$ and $\bB$ individually, we obtain $[G:H]$ equivalent 2BGA codes of length $2|H|$. \qed

Let us denote by $\tilde{\cC}$ the 2BGA code generated by the matrices $(\bA_{x,y})_{x, y \in H  }$ and $(\bB_{x,y})_{x,y \in H}$, the restriction of the matrices $\bA$ and $\bB$ to their coordinates in $H$. These matrices are non-zero since $\bA_{e, e} = \bB_{e,e} = 1$. The parameters of $\tilde{\cC}$ are related to those of $\cC$ in following way: $\ell(\cC) = [G:H] \ell(\tilde{\cC}), \, \dim(\cC) = [G:H] \dim(\tilde{\cC})$ and $d(\cC) = d(\tilde{\cC})$. Moreover, the weight of the stabilizer generators of $\tilde{\cC}$ is the same as that of the original code. Hence if the bound of Theorem~\ref{thm:quantum upper bound} holds for a subcode $\tilde{\cC}$, then it also holds for the code $\cC$.
\vspace{0.15cm}

Without loss of generality, we may thus assume that the Abelian 2BGA code $\cC$ cannot be decomposed into a direct sum of smaller Abelian 2BGA codes. Under this assumption, the isomorphism $\Bar{\Psi}^{-1}$ allows one to identify the rows of the matrices $\bH_X$ and $\bH_Z$ with vertices of $\Z^{r + s }/ \ker \Psi $. Using this identification, each physical qubit (i.e column of $ \bH_X$ or of $\bH_Z$) can be interpreted as a subset of $\Z^{r+s}/ \ker \Psi $. To apply the Bravyi-Terhal argument, one needs to index each qubit by a vertex and not a set of vertices. We thus represent the $j$-th and $(n+j)$-th qubits by $\bar{\Psi}^{-1}(g_j)$ for $j \in [[0, n-1]]$. That is to say, there are two qubits per vertex of $\Z^{r+s} / \ker(\Psi)$.

\vspace{0.1cm}
In order to apply Theorem~\ref{thm:BT bound}, we have to show that the Abelian 2BGA code $\cC$ is geometrically local, meaning that each stabilizer generator acts only on neighbouring qubits in $ \Z^{r+s}/ \ker \Psi$. The stabilizer generators are given by the rows of $\bH_X$ and $\bH_Z$ and their supports are represented respectively by 
\begin{align*}
S^X_i &= \{\bar{\Psi}^{-1}(g_j) \, | \, j \in [[ 0, n - 1 ]] \text{ s.t }  (\bH_X)_{i,j} = 1 \text{ or } (\bH_X)_{i,n+j} = 1\}, \\
S^Z_i &= \{\bar{\Psi}^{-1}(g_j) \, | \, j \in [[ 0, n - 1 ]] \text{ s.t } (\bH_Z)_{i,j} = 1 \text{ or }  (\bH_Z)_{i,n+j} = 1\}.
\end{align*}

\begin{lemma}
\label{lem:geometric locality}
Any non-trivial Abelian 2BGA code is geometrically-local. More precisely, the support of each stabilizer generator is included in a closed ball of radius $1$, for the induced metric on the quotient $\Z^{r+s} / \ker \Psi$. 
\end{lemma}

\vspace{0.15cm}
\textit{Proof:} 
For $i \in [[0,n-1]]$, the support of the $i$-th $X$-generator is $S^X_i$ which is indexed by the $i$-th row of $\bA$ and $\bB$. From the definition of the permutation matrices $\B(g)$, the support $S^X_i$ is thus given as the image of $\bar{\Psi}^{-1}$ of the two following sets:

\begin{itemize}
\item $\{g_j \, | j \in [[  0,  n - 1 ]], (\bH_X)_{i,j} = 1\} =  \{g_i, g_{a_1}^{-1} g_i, \ldots, g_{a_r}^{-1}g_i\}$,
\item $\{g_j | j \in [[  0,  n - 1 ]], \, (\bH_X)_{i,n+j} = 1\} = \{g_i, g_{b_1}^{-1}g_i, \ldots, g_{b_s}^{-1}g_i\}$. 
\end{itemize}

Hence, it follows that $S^X_i  = \{\bar{\Psi}^{-1}(g_i), \bar{\Psi}^{-1}(g_i) - \epsilon_1, \ldots, \bar{\Psi}^{-1}(g_i) - \epsilon_{r+s}\})$ which is included in $B(\bar{\Psi}^{-1}(g_i),1)$. Similarly, the support of the $i$-th $Z$-generator is $S^Z_i$ which is indexed by the $i$-th column of $\bB$ and $\bA$. From the definition of the permutation matrices $\B(g)$, the support $S^Z_i$ is thus given as the image of $\bar{\Psi}^{-1}$ of the two following sets:

\begin{itemize}
\item $\{g_j | j \in [[  0,  n - 1 ]], \,  (\bH_Z)_{i,j} = 1\} =  \{g_i, g_{b_1}g_i, \ldots, g_{b_s}g_i\}$,
\item $\{g_j | j \in [[  0,  n - 1 ]], \, (\bH_Z)_{i,n+j} = 1\} = \{g_i, g_{a_1}g_i, \ldots, g_{a_r}g_i\}$.
\end{itemize}

Hence, it follows that $S^Z_i  = \{\bar{\Psi}^{-1}(g_i), \bar{\Psi}^{-1}(g_i) + \epsilon_1, \ldots, \bar{\Psi}^{-1}(g_i) + \epsilon_{r+s}\})$ which is included $B(\bar{\Psi}^{-1}(g_i),1)$. 

\qed

\vspace{0.15cm}
We now apply our variant of the Bravyi-Terhal theorem to conclude.

\vspace{0.15cm}
\textit{Proof of Theorem~\ref{thm:quantum upper bound}:}
Let $\cC$ be a non-trivial Abelian 2BGA code of length $2n$ and with stabilizer generators parity-check matrix of weight $w \geq 4$. We have shown that without loss of generality, we may assume that the $\cC$ cannot be decomposed into a direct sum of smaller Abelian 2BGA codes and that there exists a $D$-dimensional sublattice $\Lambda$ of $\Z^D$ such that the code $\cC$ is equivalent to an Abelian 2BGA code whose qubits can be indexed by pairs on the vertices of $\Z^D / \Lambda $ with $ D = w  - 2$. By Lemma~\ref{lem:geometric locality}, the support of any stabilizer generator may be contained in a Euclidean ball of $\Z^D / \Lambda $ of radius $1$. Whenever $n^{1/D} \geq 8\sqrt{\gamma_D}$, we can apply Theorem~\ref{thm:BT bound} with $m=2$ and $\rho=1$ to conclude that the minimum distance of $\cC$ is bounded from above by $2 \sqrt{\gamma_D}(\sqrt{D} + 4)n^\frac{D-1}{D}$, where $\gamma_D$ is the Hermite constant. \qed


\section*{Acknowledgements}
All authors acknowledge the support from the Plan France 2030 through the project NISQ2LSQ, ANR-22-PETQ-0006. 


\newcommand{\etalchar}[1]{$^{#1}$}

\end{document}